\documentclass{PoS}

\usepackage{epsfig}

\let\a=\alpha      \let\g=\gamma   
\let\e=\epsilon

\newcommand{\be}{\begin{equation}}
\newcommand{\ee}{\end{equation}}
\newcommand{\bea}{\begin{eqnarray}}
\newcommand{\eea}{\end{eqnarray}}
\newcommand{\ba}{\begin{array}}
\newcommand{\ea}{\end{array}}
\def\bz{{\bar z}}
\def\tr{{\rm tr}}

\newcommand{\eq}[1]{Eq.~(\ref{#1})}
\newcommand{\fig}[1]{Fig.~\ref{#1}}

\def\Oto{\stackrel{\Omega}{\longrightarrow}}
\def\SO{S^{\rm orb}_{\rm W}}

\def\g0p{g_0^\prime}

\def\maG{\mathcal{G}}
\def\maR{\mathcal{R}}
\def\maH{\mathcal{H}}

\PoS{PoS(LAT2005)280}

\title{Gauge theories on a five-dimensional orbifold\thanks{Preprint: HU-EP-05/49}}

\ShortTitle{Gauge theories on a five-dimensional orbifold }

\author{\speaker{Francesco Knechtli} and Burkhard Bunk\\

        Institut f\"ur Physik, Humboldt Universit\"at,\\ 
        Newtonstr. 15, 12489 Berlin, Germany\\

        E-mail: \email{knechtli@physik.hu-berlin.de},
                \email{bunk@physik.hu-berlin.de}}

\author{Nikos Irges\\

        High Energy and Elementary Particle Physics Division,\\
        Department of Physics, University of Crete, 71003 Heraklion, Greece\\

        E-mail: \email{irges@physics.uoc.gr}}

\abstract{
We present a construction of non--Abelian gauge theories on the
$\mathbb{R}^4\times S^1/\mathbb{Z}_2$ orbifold. We show that no 
divergent boundary mass term for the Higgs field, identified with 
some of the fifth dimensional components of the gauge field, is generated. 
The formulation of the theories on the lattice requires only Dirichlet 
boundary conditions that specify the breaking of the gauge group. 
The first simulations in order to resolve the issue whether these 
theories can be used at low energy
as weakly interacting effective theories have been performed.
In case of a positive answer, these theories could provide us with a new
framework for studying electroweak symmetry breaking.
}

\FullConference{XXIIIrd International Symposium on Lattice Field Theory\\

                 25-30 July 2005\\

                 Trinity College, Dublin, Ireland}

\begin{document}

\section{Five--dimensional $SU(N)$ gauge theories}

The motivation to look at gauge theories in four plus one compact extra
dimensions come from an
alternative mechanism for electroweak symmetry breaking where the Higgs field is
identified with (some of) the fifth dimensional components of the gauge
field. Due to the dimensionful coupling $\g0p$, five--dimensional gauge
theories are nonrenormalizable. Nevertheless they can be employed as effective
theories at finite cutoff $\Lambda$. The claim is that the mass of the Higgs field
is finite to all orders in perturbation theory.
So far phenomenological applications of these ideas are mostly based on 1--loop
computations and we would like to understand if they are viable beyond that.

\section{A geometrical construction of the $S^1/\mathbb{Z}_2$ orbifold}

\begin{figure} \centering
\epsfig{file=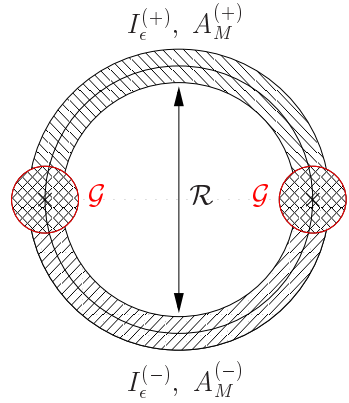, width=.35\textwidth}
\caption{Geometrical construction of the orbifold for gauge fields.}
\label{f_cha}
\end{figure}
We consider the Euclidean manifold $\mathbb{R}^4\times S^1$
parameterized by the coordinates $z=(x_\mu,x_5)$, where
$\mu=0,1,2,3$ and $x_5\in(-\pi R,\pi R]$. 
The coordinates $x_5$ are identified modulo $2\pi R$.
The definition of gauge fields on such manifold requires (at least) two open
overlapping charts. We choose
$I_\e^{(+)}=\{x_\mu,x_5\in(-\e,\pi R+\e)\}$ and
$I_\e^{(-)}=\{x_\mu,x_5\in(-\pi R-\e,\e)\}$ and denote by
$A_M^{(+)}$ and $A_M^{(-)}$ ($M=0,1,2,3,5$ is the five--dimensional Euclidean
index) the corresponding gauge fields in the Lie algebra.
The parameter $\epsilon$ determines the size of the overlaps among the charts
$O_1\,=\,\{x_\mu,x_5\in(-\e,\e)\}$ and
$O_2\,=\,\{x_\mu,x_5\in(\pi R-\e,\pi R+\e)\}$.
The situation is schematically represented in \fig{f_cha}.
In order to ensure gauge invariance, the two gauge fields on the overlaps,
where they are both defined, are related by a transition function 
$\maG(z)\in SU(N)$
\bea
 A_M^{(-)} & = & {\cal G} A_M^{(+)} {\cal G}^{-1}
                +{\cal G} \partial _M {\cal G}^{-1} 
 \,, \quad \mbox{on $O_i$, $i=1,2$} \,. \label{transition}
\eea
The orbifold projection of $S^1$ onto $S^1/\mathbb{Z}_2$ amounts to the
identification of points and fields under the $\mathbb{Z}_2$ reflection
$\maR$, which is defined to act on coordinates and fields through
\bea
 \maR\,z & = & \bz\,,\qquad \bz=(x_\mu,-x_5) \,,\\
 \maR\,A_M(z) & = & \a_MA_M(\bz) \,,\qquad
 \a_{\mu}=1\,,\quad \a_5=-1 \,.
\eea
The orbifold projection for the gauge field identifies
\bea
 \maR\,A_M^{(+)} & \equiv & A_M^{(-)} \label{orbifold}
\eea
on the overlaps, where $\maR\,A_M^{(+)}$ is defined. Outside the overlaps
the fields $A_M^{(+)}(z)$ and $A_M^{(-)}(\bz)$ are identified.
At this point we need only one gauge field, which we take to be
$A_M\equiv A_M^{(+)}$.
\eq{orbifold} together with \eq{transition} imply the {\em constraint}
\bea
 {\cal R}\,A_M & = & {\cal G}\,A_M\,{\cal G}^{-1} +
 {\cal G}\, \partial_M{\cal G}^{-1} \,, \label{constraint}
\eea
which is self--consistent when
\bea
 (\maR\,\maG)\,\maG & = & \exp(i2\pi k/N)\times\,\mathbf{1}_N
 \,,\quad k=0,1,\ldots,N-1 \,. \label{gluing}
\eea
Covariance of the constraint \eq{constraint} requires under a gauge
transformation $\Omega$
\bea
 {\cal G} & \Oto & ({\cal R}\,\Omega)\,{\cal G}\,\Omega^{-1} \,. \label{gtrG}
\eea
The gauge covariant derivative of $\maG$ is then defined through
\bea
 D_M {\cal G} & = & \partial_M {\cal G}
 + ({\cal R}\,A_M) \, {\cal G} - {\cal G} \, A_M \;\equiv\; 0 \label{covder}
\eea
and by virtue of the constraint \eq{constraint} it vanishes identically.

The fundamental domain of the orbifold is the strip
$I_0=\{x_\mu,x_5\in[0,\pi R]\}$.
The gauge theory on $I_0$ is obtained by starting from the gauge invariant theory
formulated on the chart $I_\e\equiv I_\e^{(+)}=\{x_\mu,x_5\in(-\e,\pi R+\e)\}$ in
terms of the gauge field $A_M$ and the {\em spurion} field (the transition function)
$\maG$. This theory is gauge invariant under gauge transformations that obey
\bea
 \maR\,\Omega & = & \Omega \,.
\eea
The spurion field transforms like the field strength tensor, see \eq{gtrG}.
The parameter $\e$ is then set to zero and the spurion field subject to the
{\em boundary conditions}
\bea
 {\cal G}(x_\mu,x_5=0) \,= & g & =\, {\cal G}(x_\mu,x_5=\pi R) \,, \label{bcG}
\eea
where $g$ is a constant matrix. It follows from \eq{gluing} that $g^2$ is an
element of the center of $SU(N)$.
At the boundaries $x_5=0$ and $x_5=\pi R$ of the strip all
derivatives of $\maG$ are required to vanish. From \eq{constraint} the boundary
conditions for {\em any} field can be derived, for example
\bea
  \alpha_M\,A_M \, = \, g\,A_M\,g^{-1}
  & & \mbox{Dirichlet boundary conditions} \,, \label{Dbc} \\
  -\a_M\,\partial_5A_M \, = \, g\,\partial_5A_M\,g^{-1}
  & & \mbox{Neumann boundary conditions} \,. \label{Nbc}
\eea
It is clear from \eq{gtrG} that only the gauge transformations satisfying
\bea
 [g,\Omega(z)] & = & 0 \,,\quad {\rm at}\;x_5=0\;{\rm and}\;x_5=\pi R
 \label{breaking}
\eea
are still a symmetry: the gauge group is {\em broken} at the boundaries if $g\neq1$.

A parameterization of the matrix $g$ is given by
\bea
 g & = & {\rm e}^{-2\pi iV\cdot H} \label{gma}
\eea
where $H=\{H_i\}$, $i=1,\ldots,N-1$ are the Hermitian generators of the
Cartan subalgebra of $SU(N)$ and $V=\{V_i\}$ is the twist vector of the orbifold.
It follows that under group conjugation by $g$ the Hermitian generators $T^A$ of
$SU(N)$ transform as
$g\,T^A\,g^{-1} = \eta^A\,T^A$, $\eta^A=\pm1$ \cite{NF}.
The breaking of the gauge group in \eq{breaking} is determined by the choice
of the twist vector and is of the form
\bea
 SU(p+q) & \longrightarrow & \maH=SU(p)\times SU(q)\times U(1) \,.
\eea
\eq{Dbc} means that only the components $A_5^A$ associated with generators $T^A$,
which do not commute with $g$ survive at the orbifold boundaries.
Therefore we identify $[A_5(z),g]$
with the {\em Higgs field}. It transforms in some representation of $\maH$.

It is plausible that we can apply the Symanzik analysis of counterterms
for renormalizable field theories with boundaries \cite{Sy,LuSy} to the
nonrenormalizable orbifold theory defined on the strip $I_0$. The latter
is considered as effective theory at finite cutoff $\Lambda$.
The boundary term
\bea
 {\rm \tr}\,\{[A_5,g][A_5,g]\} \label{higgsbm}
\eea
is a Higgs mass term and is invariant under gauge transformations of
the unbroken subgroup $\maH$. It would be a quadratically divergent (with the
cutoff $\Lambda$) boundary mass term. Explicit 1--loop calculations \cite{NQ1}
indicate that this term is not present and a shift symmetry argument forbids it
\cite{NQ2}.

In our geometrical construction the term \eq{higgsbm} has to be derived from
a {\em gauge invariant} term in the theory formulated on the chart $I_\e$.
It could come from
\bea
 {\rm tr}\,\{D_5{\cal G}\,D_5{\cal G}\} & \equiv & 0
\eea
but this term vanishes identically due to \eq{covder}. In fact there are no
boundary terms of dimension less than or equal to four. The lowest dimensional
boundary terms are the dimension five terms
\bea
 \frac{1}{{\g0p}^2}\,{\rm Re}\,{\rm tr}\,\{g\,F_{MN}\,F_{MN}\} & \mbox{and} &
 \frac{1}{{\g0p}^2}\,{\rm Re}\,{\rm tr}\,\{g\,F_{MN}\,g\,F_{MN}\} \,.
\eea

\section{Lattice simulations of $SU(2)$}

On the lattice we define the orbifold theory
in the strip $I_0=\{z=a(n_\mu,n_5)|\;0\le n_5\le N_5=\frac{\pi R}{a}\}$,
where $a$ is the lattice spacing. In the following we assume periodic boundary
conditions in the directions $\mu=0,1,2,3$. The Wilson action for the orbifold
reads
\bea
 \SO[U] \,=\, \frac{\beta}{2N}\sum_pw(p)\,\tr\{1-U(p)\}\,,\quad
 w(p) \,=\, \left\{\begin{array}{ll}
                   \frac{1}{2} & \mbox{$p$ in the boundary} \\
                   1 & \mbox{in all other cases.}
                   \end{array} \right. \label{orbia}
\eea
where the sum is over oriented plaquettes $p$. The bare dimensionful gauge coupling
$\g0p$ on the lattice is defined through $\beta=2Na/{\g0p}^2$. The action
\eq{orbia} is identical to the one for the Schr\"odinger Functional (SF)
\cite{LNWW}.

As it was shown in Ref. \cite{NF}, on the lattice the orbifold is specified by
imposing the following Dirichlet boundary conditions in the four--dimensional
boundary planes of the strip $I_0$
\bea
 U(z,\mu) & = & g\,U(z,\mu)\,g^{-1} \,,\quad \mbox{at $n_5=0$ and $n_5=N_5$} \,.
 \label{Dbclat}
\eea
We emphasize that these boundary conditions are of a different type than for the
SF, as they {\em constrain} the gauge variables at the boundaries but do not fix
them completely. In the naive continuum limit the Dirichlet and Neumann boundary
conditions, the latter ``carried'' by the gluon propagator, are recovered \cite{NF}.

We present here first results from simulations of the five--dimensional
$SU(2)$ gauge theory on the orbifold. The matrix $g$ in \eq{gma} is given by
$g=-i\sigma_3$ ($V=1/2$) and \eq{Dbclat} implies that the boundary links
are parameterized in terms of a $U(1)$ phase $\phi(z,\mu)$
\bea
 U(z,\mu) & = & \exp(i\phi(z,\mu)\sigma_3)
 \,,\quad \mbox{at $n_5=0$ and $n_5=N_5$} \,. \label{u1links}
\eea
The simulation algorithm is composed of heatbath and overrelaxation updates
in the bulk, where the gauge group is $SU(2)$, and for the four--dimensional
$U(1)$ boundary links. 
%
%
%
\begin{figure} \centering
\epsfig{file=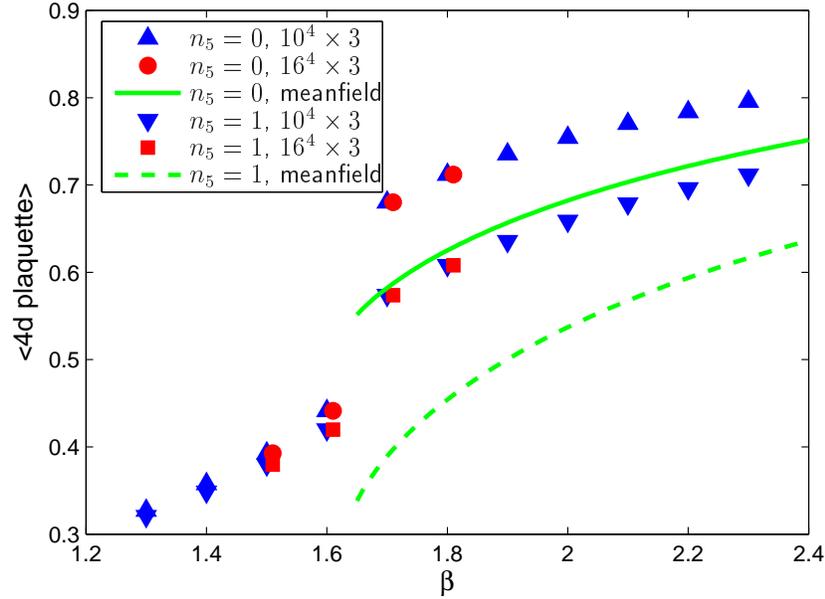, width=.75\textwidth}
\caption{Orbifold boundaries and bulk.}
\label{f_bborbi}
\end{figure}

Five--dimensional $SU(N)$ gauge theories in infinite volume have a phase transition
separating the Coulomb phase with massless gluons at very large $\beta$ from the
confining phase at very small $\beta$ \cite{Cr,Wi}. For $SU(2)$ the phase transition
is at $\beta_c=1.642(15)$ \cite{Cr}. In \fig{f_bborbi} we show results of
simulations of the orbifold with $N_5=3$ and two different four--dimensional
volumes $10^4$ and $16^4$. We plot the expectation values of the four--dimensional
plaquettes at $n_5=0$ and $n_5=1$ (equal to the ones at $n_5=3$ and $n_5=2$
respectively). We see that the orbifold is similar to the torus geometry
in the sense that around $\beta=1.6$ there is a jump of plaquette values and we
observed hysteresis effects. In the middle of the extra dimension the plaquette
values are very close to the ones on a $10^4\times4$ torus. At the boundaries the
plaquettes are ``colder''. We do not observe significant effects when changing the
four--dimensional volume.

We have done a meanfield computation for the orbifold geometry as follows.
We set
\bea
 U(z,\mu) & = & u(n_5)\,\times\,\mathbf{1}_2\,, \qquad\qquad\quad
 n_5\,=\,0,\ldots,N_5 \\
 U(z,5)   & = & u(n_5+1/2)\,\times\,\mathbf{1}_2\,, \qquad 
 n_5\,=\,0,\ldots,N_5-1
\eea
The meanfield computation amounts to an iterative solution of a system of equations
for the factors $u$. Each link is equated to its expectation value in the
fixed configuration of all the other links for a given value of $\beta$.
We get,
for the boundary links at $n_5=0$ (plus sign) and $n_5=N_5$ (minus sign):
\bea
 b  & = & \beta [(1/2)\,6\,u(n_5)^3 + u(n_5\pm1)\,u(n_5\pm1/2)^2] \\
 u(n_5) & = & I_1(b)/I_0(b) \,,
\eea
for the $SU(2)$--links $U(z,\mu)$ in the bulk at $n_5=1,\ldots,(N_5-1)$:
\bea
 b & = & \beta [6\,u(n_5)^3 + u(n_5+1/2)^2\,u(n_5+1) + u(n_5-1/2)^2\,u(n_5-1)] \\
 u(n_5) & = & I_2(b)/I_1(b) \,,
\eea
and for the $SU(2)$--links $U(z,5)$ along the extra dimension
at $n_5=(1/2)\ldots(N_5-1/2)$
\bea
 b & = & \beta\,8\,u(n_5-1/2)\,u(n_5)\,u(n_5+1/2) \\
 u(n_5) & = & I_2(b)/I_1(b) \,.
\eea
The meanfield plaquette values $u(0)^4$ and $u(1)^4$ for $N_5=3$ are plotted
in \fig{f_bborbi}
for ``large'' $\beta$ values (at lower $\beta$ only the solution $u=0$ is found).
They show qualitatively the difference between plaquettes at the boundaries and in
the middle of the extra dimension as it is seen in the simulations.


\begin{thebibliography}{99}

\bibitem{NF} N.~Irges and F.~Knechtli, \emph{Nonperturbative definition of
             five--dimensional gauge theories on the
             $\mathbb{R}^4\times S^1/\mathbb{Z}_2$ orbifold},
             \emph{Nucl. Phys.} {\bf B719} (2005) 121,
             [{\tt hep-lat/0411018}].
\bibitem{Sy} K.~Symanzik, \emph{Schr\"odinger representation and Casimir effect
             in renormalizable quantum field theory},
             \emph{Nucl. Phys.} {\bf B190} (1981) 1.
\bibitem{LuSy} M.~L\"uscher, \emph{Schr\"odinger representation in quantum field
               theory},
               \emph{Nucl. Phys.} {\bf B254} (1985) 52.
\bibitem{NQ1} G.~von Gersdorff, N.~Irges and M.~Quiros, \emph{Bulk and brane
              radiative effects in gauge theories on orbifolds},
              \emph{Nucl. Phys.} {\bf B635} (2002) 127,
              [{\tt hep-th/0204223}].
\bibitem{NQ2} G.~von Gersdorff, N.~Irges and M.~Quiros, \emph{Radiative
              brane--mass terms in $D>5$ orbifold gauge theories},
              \emph{Phys. Lett.} {\bf B551} (2003) 351,
              [{\tt hep-ph/0210134}].
\bibitem{LNWW} M.~L\"uscher, R.~Narayanan, P.~Weisz and U.~Wolff,
               \emph{The Schr\"odinger Functional: a renormalizable probe for
               non--Abelian gauge theories}, 
               \emph{Nucl. Phys.} {\bf B384} (1992) 168,
               [{\tt hep-lat/9207009}].
\bibitem{Cr} M.~Creutz, \emph{Confinement and the critical dimensionality of
             space--time},
             \emph{Phys. Rev. Lett.} {\bf 43} (1979) 553.
\bibitem{Wi} B.B.~Beard, R.C.~Brower, S.~Chandrasekharan, D.~Chen, A.~Tsapalis and
             U.-J.~Wiese, \emph{D--theory: field theory via dimensional reduction
             of discrete variables},
             \emph{Nucl. Phys.} {\bf B(Proc. Suppl.)63} (1998) 775,
             [{\tt hep-lat/9709120}].

\end{thebibliography}
\end{document}